\documentclass[10pt,twocolumn]{article}

\usepackage[dvips]{graphicx}
\usepackage[english]{babel}

\usepackage{amsmath}

\usepackage{slashed}

\allowdisplaybreaks[1]

\newcommand{\half}{{\textstyle\frac{1}{2}}}

\newcommand{\gev}{\operatorname{GeV}}
\newcommand{\mev}{\operatorname{MeV}}

\newcommand{\ms}{\mskip 1.5mu}
\newcommand{\bs}{\mskip -1.5mu}

\newcommand{\lrD}{{D^{\hspace{-0.8em}%
      \raisebox{0.8ex}{$\scriptstyle\leftrightarrow$}}}{}}

\newcommand{\lrpartial}{{\partial^{\hspace{-0.65em}%
      \raisebox{0.8ex}{$\scriptstyle\leftrightarrow$}}}\hspace{-0.15em}{}}
\newcommand{\lpartial}{{\partial^{\hspace{-0.65em}%
      \raisebox{0.8ex}{$\scriptstyle\leftarrow$}}}\hspace{-0.15em}{}}
\newcommand{\rpartial}{{\partial^{\hspace{-0.65em}%
      \raisebox{0.8ex}{$\scriptstyle\rightarrow$}}}\hspace{-0.15em}{}}

\newcommand{\tvec}[1]{\boldsymbol{#1}}

\begin{document}

\title{%
\raggedleft{\normalsize CPHT-RR016.0310, \
DESY 10-041, LPT-Orsay-10-21} \\[1.0em]
\centering\textbf{\Large The transverse spin structure of the pion
  at short distances}}

\author{Markus~Diehl \\[0.2em]
\textit{Deutsches Elektronen-Synchroton DESY, 22603 Hamburg, Germany}
\and
Lech~Szymanowski \\[0.2em]
\textit{Soltan Institute for Nuclear Studies, Warsaw, Poland \&} \\
\textit{Centre de Physique Th\'eorique, \'Ecole Polytechnique, CNRS, 91128
  Palaiseau, France \&} \\
\textit{Laboratoire de Physique Th\'eorique, Universit\'e Paris-Sud, CNRS,
  91405 Orsay, France}
}

\date{\parbox{0.9\textwidth}{\normalsize \textbf{Abstract:} We study the
    form factors of the quark tensor currents in the pion at large squared
    momentum transfer $Q^2$.  It turns out that certain form factors can
    be evaluated using collinear factorization, whereas others receive
    important contributions from the end-point regions of the longitudinal
    quark momenta in the pion.  We derive simple analytic expressions for
    the dominant terms at high $Q^2$ and illustrate them numerically.
}}

\maketitle

%\linenumbers

%%%%%%%%%%%%%%%%%%%%%%%%%%%%%%%%%%%%%%%%%%%%%%%%%%%%%%%%%%%%%

\section{Introduction}
\label{sec:intro}

The structure of the pion at short distances unites two characteristic
features of quantum chromodynamics.  On the one hand, the pion plays a
unique role among hadrons as the Goldstone boson of spontaneous chiral
symmetry breaking.  On the other hand, asymptotic freedom is central for
understanding its structure at short distances, where quarks and gluons
interact perturbatively as in any other hadron.  Moreover, many studies of
hadron structure are very much simplified when one deals with spin-zero
hadrons, and the pion is probably the spin-zero hadron for which most
quantitative information is available, both from experiment and from
calculations in lattice QCD.  A versatile tool to describe hadronic
structure is given by generalized parton distributions or, equivalently,
by the form factors of a tower of local quark-gluon operators containing
an increasing number of covariant derivatives.

A perhaps surprising feature of the pion is that is has a non-trivial spin
structure.  An instructive quantity to describe this structure is the
distribution $\rho(x,\tvec{b})$ of quarks with longitudinal momentum
fraction $x$ and transverse distance $\tvec{b}$ from the center of the
pion \cite{Burkardt:2002hr}.  Due to parity invariance this distribution
cannot depend on the longitudinal quark polarization.  However, the
distribution of quarks with transverse spin $\tvec{s}$ has a polarization
dependent part, which is proportional to $(\tvec{s} \times \tvec{b})^z$
and was found to be sizeable in a recent lattice study
\cite{Brommel:2007xd}.
This polarization dependence can be quantified by the form factors of the
quark tensor operator $\bar{q}\ms i\sigma^{\alpha\beta}\ms q$ and its
analogs containing covariant derivatives.  The present work is concerned
with these tensor form factors at high momentum transfer, or in other
words with the correlation between the transverse polarization and the
transverse position of quarks very close to the center of the pion.

Form factors at high momentum transfer have played a key role in the early
development of methods for calculating exclusive observables in QCD
\cite{Efremov:1979qk,Lepage:1979zb}.  They continue to provide an
important area for applying factorization, with close links to the physics
of exclusive $B$ meson decays.  In the limit of infinite momentum transfer
$Q^2$ form factors can be described within standard collinear
factorization, but extensive studies of the electromagnetic pion form
factor $F_\pi(Q^2)$ indicate that at experimentally accessible values of
$Q^2$ this description receives important corrections, see for instance
\cite{Nesterenko:1982gc,Li:1992nu,Jakob:1993iw,Braun:1999uj,Bakulev:2004cu}.
In the present work we aim at providing a baseline for the large $Q^2$
behavior of the pion tensor form factors $B_{T ni}$, and we will use a
very simplified extension of the collinear factorization framework that
allows us to obtain expressions in compact analytical form.  We do
therefore not expect our results to be quantitatively reliable at
moderately large $Q^2$, and we will in particular refrain from comparing
to the lattice calculations in \cite{Brommel:2007xd}, which go up to $Q^2
\approx 2.5 \gev^2$.  On the other hand, our analytic expressions may be
of use if one wants to devise parameterizations of $B_{T ni}(Q^2)$ that
have the correct behavior at large $Q^2$.

The large $Q^2$ behavior of pion tensor form factors is also interesting
because it involves pion distribution amplitudes of twist three, which
have a particular behavior at the end-points of the momentum fraction
variable \cite{Braun:1989iv}.  We find that for certain form factors $B_{T
  ni}$ the formulae obtained by using collinear factorization have
end-point divergences and hence need to be modified.  This is similar to
other cases where twist-three pion distribution amplitudes appear, such as
spectator interactions in exclusive $B$ decays
\cite{Beneke:2000wa,Beneke:2001ev}, pion electroproduction $ep\to e \pi^+
n$ with transverse polarization of the exchanged virtual photon
\cite{Goloskokov:2009ia}, and certain power corrections to $F_\pi(Q^2)$
\cite{Geshkenbein:1982zs,Geshkenbein:1984jh}.

This paper is organized as follows.  In the next section we set up the
calculational framework used in the present work.  In
Sect.~\ref{sec:extract} we extract the contributions from the
hard-scattering graphs that dominate in the large $Q^2$ limit and derive
simple analytic expressions for the form factors $B_{T ni}$.  In
Sect.~\ref{sec:numerical} we present some numerical illustrations of our
results, and in Sect.~\ref{sec:summary} we summarize our findings.

%%%%%%%%%%%%%%%%%%%%%%%%%%%%%%%%%%%%%%%%%%%%%%%%%%%%%%%%%%%%%

\section{Setting up the calculation}
\label{sec:setup}

The tensor form factors of the pion parameterize the matrix elements of
the local operators
\begin{equation}
  \label{tensor-op}
\operatorname{T}
\underset{(\alpha,\beta_1)}{\operatorname{A}} \;
\underset{(\beta_1,\ldots,\beta_n)}{\operatorname{S}}
\bar{q}\, i\sigma^{\alpha\beta_1}\,
   i\lrD^{\beta_2} \cdots\, i\lrD^{\beta_n}\ms q \, ,
\end{equation}
where $\lrD^\beta = \lrpartial^\beta - i g A^\beta$ with $\lrpartial^\beta
= \half (\rpartial^\beta - \lpartial^\beta)$ is the covariant derivative.
Here $S$ and $A$ respectively denote symmetrization and antisymmetrization
in the indicated indices, and $T$ denotes the subtraction of traces in all
index pairs.  These operations, which project on operators with twist two,
can be implemented in a simple way by contraction with two constant
auxiliary vectors $a$, $b$ satisfying $a^2 = ab = 0$ and $b^2 \neq 0$
\cite{Diehl:2006js}.  The tensor form factors are then given by\footnote{%
  A factor $i$ is missing on the r.h.s.\ of eq.~(71) in
  \protect\cite{Diehl:2006js}.}
\begin{align}
  \label{ff-def}
& \langle \pi^+(p') \,\big|\, \bar{u}
  \ms i\sigma^{\alpha\beta} a_\alpha b_\beta\,
  (i\lrD a)^{n-1}\ms u \,\big|\, \pi^+(p) \rangle
 = (aP)^{n-1} 
\nonumber \\
&\quad \times
  \frac{(ap) (bp') - (bp) (ap')}{m_\pi}\, 
  \sum_{\genfrac{}{}{0pt}{}{i=0}{\text{even}}}^{n-1}
    (2\xi)^i\ms B_{T ni}^{u}(Q^2)
\end{align}
with $Q^2 = - (p-p)^2$ and
\begin{align}
P &= \frac{1}{2} (p+p') , &
\xi &= \frac{a (p-p')}{a (p+p')} .
\end{align}
The form factors in \eqref{ff-def} refer to $u$-quarks; those for
$d$-quarks follow from isospin symmetry and read
\begin{align}
B^d_{T ni} &= (-1)^{n}\ms B^u_{T ni} .
\end{align}
The form factors can be written as Mellin moments of generalized parton
distributions of the pion as shown in \cite{Diehl:2006js}, but we will not
need this representation here.

\begin{figure}[t!]
\begin{center}
\includegraphics[width=0.49\textwidth]{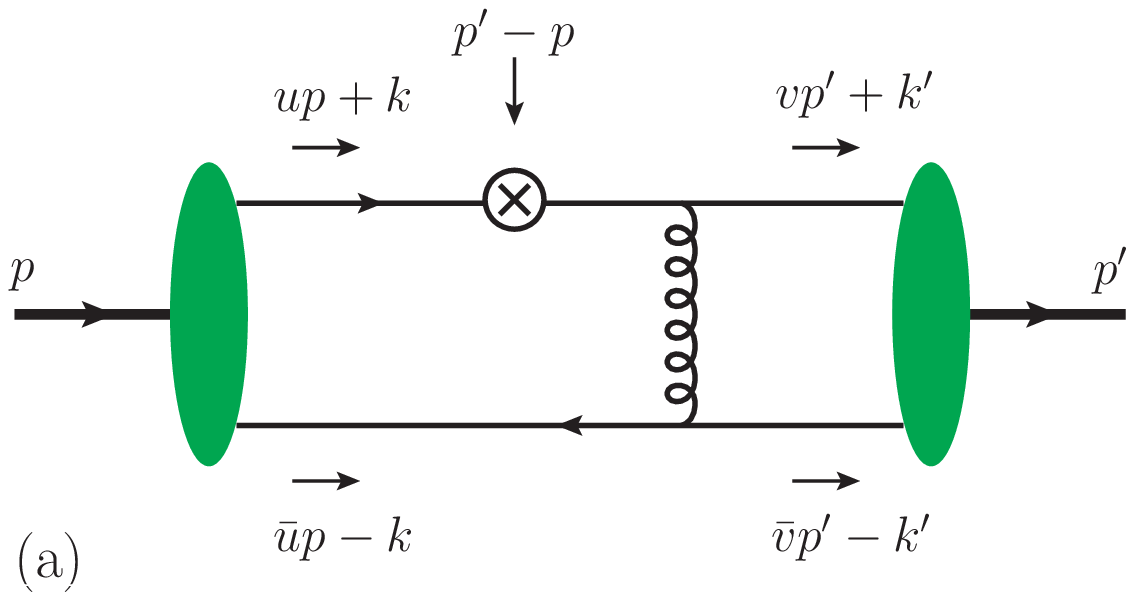} \\[1.5em]
\includegraphics[width=0.49\textwidth]{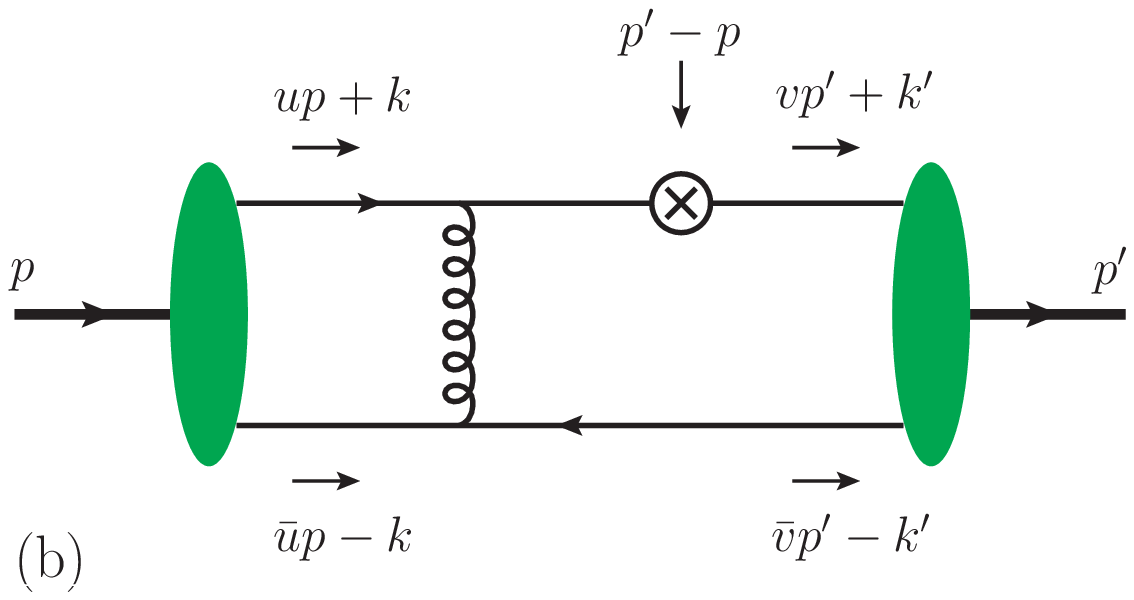} \\[1.5em]
\includegraphics[width=0.49\textwidth]{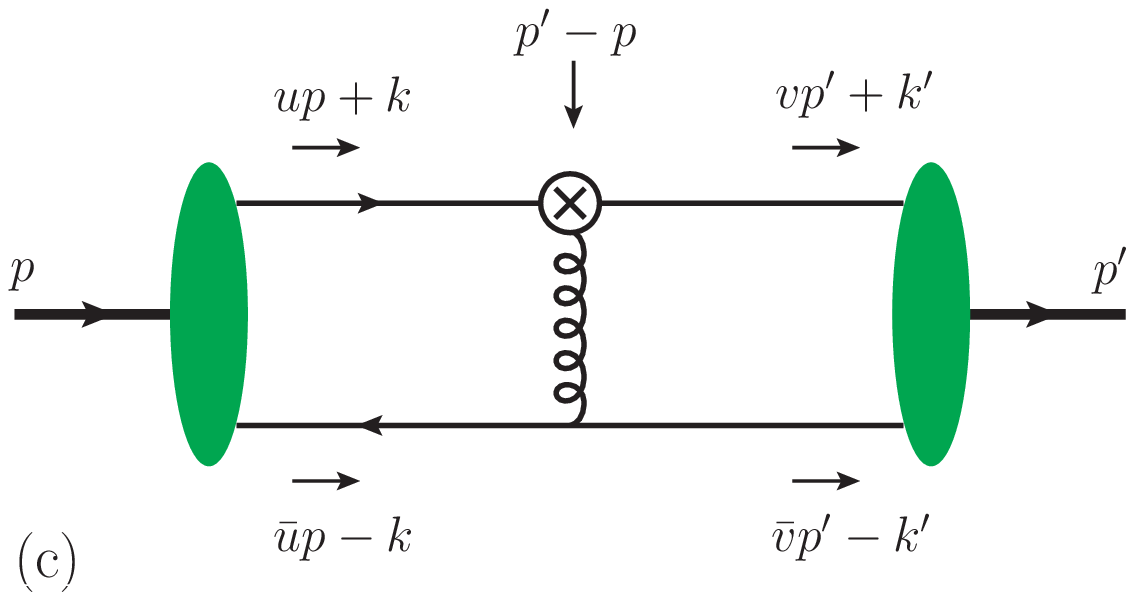}
\end{center}
\caption{\label{fig:graphs} Graphs for the matrix element
  \protect\eqref{ff-def} in the limit of large $Q^2$.  The crossed circle
  represents the insertion of the relevant current operator, given by
  \protect\eqref{tensor-no-g} for graphs (a) and (b) and by
  \protect\eqref{tensor-one-g} for graph (c).  The blobs stand for the sum
  of twist-two and twist-three distribution amplitudes as specified in
  \protect\eqref{projectors}.}
\end{figure}

In the collinear factorization formalism and at leading order in
$\alpha_s$ the matrix element \eqref{ff-def} receives contributions from
the graphs in figure~\ref{fig:graphs}.  Due to the covariant derivatives,
the operator \eqref{tensor-op} contains terms with zero to $n-1$ gluon
fields.  Graphs (a) and (b) correspond to the term without gluon fields,
i.e.\ to
\begin{equation}
  \label{tensor-no-g}
\bar{u} \ms i\sigma^{\alpha\beta} a_\alpha b_\beta\,
  (i\lrpartial a)^{n-1}\ms u
\end{equation}
in \eqref{ff-def}.  The same graphs describe the electromagnetic pion form
factor if one inserts the electromagnetic current instead of the current
in \eqref{tensor-no-g}.  Graph (c) corresponds to the terms in
\eqref{tensor-op} that have exactly one gluon field, i.e.\ to
\begin{equation}
  \label{tensor-one-g}
  \sum_{j=1}^{n-1} \bar{u} \ms i\sigma^{\alpha\beta} a_\alpha b_\beta\,
  (i\lrpartial a)^{n-1-j} (g A a) \, (i\lrpartial a)^{j-1}\ms u
\end{equation}
in \eqref{ff-def}.  Terms with more than one gluon field do not contribute
at this level.

When calculating the hard-scattering part of the graphs we neglect the
pion mass, so that the pion momenta $p$ and $p'$ are purely lightlike.  We
use them to define the two light-cone directions required for specifying
the distribution amplitudes of the pions, working in a reference frame
where the incoming pion moves in the positive and the outgoing pion in the
negative $z$ direction.  As indicated in the figure, we write the
$u$-quark momentum as $u p + k$ in the incoming $\pi^+$ and as $v p' + k'$
for the outgoing $\pi^+$, with the light-cone momentum fractions $u$ and
$v$ ranging from $0$ to $1$.  The vectors $k$ and $k'$ are transverse to
both $p$ and $p'$.  We neglect the small momentum components of the quarks
and antiquarks, i.e.\ the component along $p'$ in the incoming pion and
the component along $p$ in the outgoing one.  Note that $(up + k)^2 = k^2$
and $(vp' + k')^2 = k'^2$ are in general not zero---we will comment on
this shortly.

Since the tensor operators \eqref{tensor-op} have odd chirality, we need
one chiral-even and one chiral-odd pion distribution amplitude in the
graphs to obtain a nonvanishing hard-scattering amplitude.  Since there is
no chiral-odd pion distribution amplitude with twist two, we must go to
twist-three level.  The relevant distribution amplitudes have been
introduced in \cite{Braun:1989iv}.  After a Fourier transform from the
position representation used in \cite{Braun:1989iv} to momentum space, the
projection operators for the incoming and the outgoing pion respectively
read \cite{Beneke:2000wa}
\begin{align}
  \label{projectors}
& \Phi\Bigl( u,\frac{\partial}{\partial k} \Bigr)
 = - \frac{i f_\pi}{4} \biggl\{ \phi(u)\ms \slashed{p} \gamma_5
    + \mu_\pi\ms \phi_p(u)\ms \gamma_5
\nonumber \\[0.2em]
 & \ + \mu_\pi\ms \frac{i\sigma^{\alpha\beta} \gamma_5}{6}
      \biggl[ \frac{d\phi_\sigma(u)}{du}\,
              \frac{p_\alpha\ms p'\!\bs{}_\beta}{p p'}
            - \phi_\sigma(u)\, p_\alpha\ms
              \frac{\partial}{\partial k^\beta} \biggr] \biggr\} \, ,
\nonumber \\[0.3em]
& \Phi'\Bigl( v,\frac{\partial}{\partial k'} \Bigr)
 = \frac{i f_\pi}{4} \biggl\{ \phi(v)\ms \slashed{p}' \gamma_5
    - \mu_\pi\ms \phi_p(v)\ms \gamma_5
\nonumber \\[0.2em]
 & \ + \mu_\pi\ms \frac{i\sigma^{\alpha\beta} \gamma_5}{6}
      \biggl[ \frac{d\phi_\sigma(v)}{dv}\, 
              \frac{p'\!\bs{}_\alpha\ms p_\beta}{p p'}
            - \phi_\sigma(v)\, p'\!\bs{}_\alpha\ms
              \frac{\partial}{\partial k'^{\beta}}
      \biggr] \biggr\}
\end{align}
with $f_\pi = 130.4\mev$ \cite{Amsler:2008zzb} and
\begin{equation}
  \label{chiral-param}
  \mu_\pi = \frac{m_{\pi}^2}{m_u+m_d} .
\end{equation}
In \eqref{chiral-param} the pion mass can of course not be neglected since
one is dealing with a non-perturbative quantity.  For the twist-three
distribution amplitudes we take the asymptotic forms under evolution,
\begin{align}
  \label{twist-3-asy}
\phi_p(u) &= 1, & \phi_\sigma(u) &= 6 u\bar{u} ,
\end{align}
where here and in the following we use the notation
\begin{equation}
\bar{u} = 1-u .
\end{equation}
The normalization constant $f_{3\pi}$ associated with the twist-three
quark-gluon-quark distribution amplitudes of the pion asymptotically
evolves to zero \cite{Braun:1989iv}.  In the limit where $\phi_p$ and
$\phi_\sigma$ take the form \eqref{twist-3-asy}, the graphs in
figure~\ref{fig:graphs} therefore give the full answer for the matrix
element \eqref{ff-def}.  Conversely, the consideration of distribution
amplitudes deviating from \eqref{twist-3-asy} would require the inclusion
of graphs with an additional gluon in one of the pion distribution
amplitudes and thus considerably complicate the analysis.  Since in this
work we aim at understanding the basic behavior of the form factors at
large $Q^2$, we consider the restriction to the asymptotic forms
\eqref{twist-3-asy} to be sufficient.  On the other hand, we can easily
keep the general form
\begin{equation}
  \label{phi-def}
\phi(u) = 6 u\bar{u}\ms g(u)
\end{equation}
of the twist-two distribution amplitude, where
\begin{equation}
  \label{Gegenbauer}
g(u) = 1 + \sum_{n=2}^{\infty} a_n^{}\ms C_n^{3/2}(u-\bar{u})
\end{equation}
is the usual expansion in Gegenbauer polynomials, with coefficients $a_n$
that evolve with a simple multiplicative factor at leading order
\cite{Efremov:1979qk,Lepage:1979zb}.  With \eqref{twist-3-asy} to
\eqref{Gegenbauer} the factorization scale dependence of the projectors
\eqref{projectors} is then given by
\begin{align}
  \label{evolution}
\mu_\pi(\mu) &= \mu_\pi(\mu_0) \,
  \biggl( \frac{\alpha_s(\mu_0)}{\alpha_s(\mu)} \biggr)^{4/\beta_0} ,
\nonumber \\
a_n(\mu)     &= a_n(\mu_0) \,
  \biggl( \frac{\alpha_s(\mu_0)}{\alpha_s(\mu)} \biggr)^{-\gamma_n/\beta_0}
\end{align}
at leading logarithmic accuracy, where $\alpha_s(\mu)$ is the one-loop
running coupling, $\beta_0 = 11 - 2 n_F /3$, and the first few anomalous
dimensions read $\gamma_2 = 50/9$, $\gamma_4 = 364 /45$, etc.  The scale
dependence of $\mu_\pi$ simply reflects the running of the quark masses in
\eqref{chiral-param}.

An alternative form of the projector \eqref{projectors} was derived in
section 3.2 of \cite{Beneke:2001ev}, which had earlier been used in
\cite{Geshkenbein:1982zs,Geshkenbein:1984jh}.  This derivation requires
one to keep the small components of the quark and antiquark momenta in the
intermediate stages of the calculation and to adjust them such that both
the quark and the antiquark attached to the pion wave function are exactly
on shell.  Having the external quarks and antiquarks of the
hard-scattering subprocess exactly on shell is certainly an attractive
feature of the calculation, especially from the point of view of gauge
invariance.  It comes, however, at the price of violating momentum
conservation.  Consider for definiteness the quark and antiquark momenta
in the incoming pion:
\begin{align}
k_q &= u p + k + w_q\ms p' , &
k_{\bar{q}} &= \bar{u} p - k + w_{\bar{q}}\ms p' .
\end{align}
For generic values of $u$ and $k$ one cannot have both $k_q^2 =
k_{\bar{q}}^2 = 0$ and $w_q + w_{\bar{q}} = 0$ (for this it does not
matter whether one neglects the pion mass or not).  In our calculation, we
choose to be consistent with momentum conservation neglect the small
components $w_q\ms p'$ and $w_{\bar{q}}\ms p'$.  We will explicitly check
that gauge invariance holds for the class of covariant gauges and within
the accuracy of our calculation.

As explained in \cite{Beneke:2000wa}, the derivatives with respect to $k$
and $k'$ in the projector \eqref{projectors} act on the hard-scattering
kernel before one takes the collinear limit by setting $k=k'=0$.  However,
we will see that for some of the form factors $B_{T ni}^u$ the collinear
limit cannot be taken since the integrals over $u$ and $v$ diverge at
their end-points for $k=k'=0$.  To keep the intermediate steps of our
calculation well-defined, we introduce transverse-momentum dependent
factors $\Sigma(u,k^2)$ and $\Sigma(v,k'^2)$ for the incoming and outgoing
pion.  These factors are real-valued and normalized as
\begin{equation}
  \label{sigma-norm}
\int d^2 k\; \Sigma(u,k^2) = 1 .
\end{equation}

In a more sophisticated approach, which has for instance been used in
\cite{Goloskokov:2009ia}, one would multiply the different terms in $\Phi$
and $\Phi'$ with different factors and interpret the result as pion
light-cone wave functions that depend on both a longitudinal momentum
fraction $u$ or $v$ and on the transverse parton momentum.  Furthermore,
in the spirit of the modified hard-scattering approach, one should include
Sudakov factors for each pion, which resum a class of large logarithms
from higher-order corrections and depends on the momentum fractions, the
transverse parton momenta and the hard scale $Q$ in a non-trivial way
\cite{Li:1992nu}.  Formally, the Sudakov factors alone would already
remove the end-point divergences of the $u$ and $v$ integrals, but for a
wide range of hard scales $Q^2$ the resulting integrals will receive large
contributions from phase space regions where parton virtualities are low
and the perturbative expression of the Sudakov factors is not justified
(see \cite{DescotesGenon:2001hm} for a detailed analysis of the situation
in semileptonic $B\to \pi$ decays).  Moreover, even a calculation with
Sudakov factors but without a nonperturbative transverse-momentum
dependence of the pion wave function would not readily yield simple
analytic expressions.  Since the latter is what we are aiming for in the
present work, we will use a global factor $\Sigma(u,k^2)\, \Sigma(v,k'^2)$
as a minimal version to regulate the intermediate steps of our calculation
and simplify the resulting integrals in the end, see
eq.~\eqref{k-to-lambda} below.

With these preliminaries we can write the large-$Q^2$ limit of the matrix
element we are interested in as
\begin{align}
  \label{mat-el-initial}
& \langle \pi^+(p') \,\big|\, \bar{u} \ms
  i\sigma^{\alpha\beta} a_\alpha b_\beta\,
  (i\lrD a)^{n-1} \ms u \,\big|\, \pi^+(p) \rangle
\nonumber \\[0.2em]
& \quad = 4\pi\alpha_s \frac{C_F}{N_c}\, 6 f_\pi^2\, \mu_\pi^{}
  \int du\, dv\; d^2 k\, d^2k'\; 
   \Sigma(u,k^2)\,
\nonumber \\[0.2em]
& \hspace{9em} \times  \Sigma(v,k'^2)\, f(u,v; k, k')
\end{align}
with
\begin{align}
  \label{f-initial}
& f(u,v; k, k') =
    \frac{1}{6 f_\pi^2\, \mu_\pi^{}}\,
\nonumber \\[0.1em]
& \quad \times \operatorname{Tr}
    \Phi\Bigl( u,\frac{\partial}{\partial k} \Bigr) \,
    \gamma^\lambda \ms
    \Phi'\Bigl( v,\frac{\partial}{\partial k'} \Bigr) \,
    \frac{D_{\lambda\mu}}{\bar{u} \bar{v} Q^2 - (k - k')^2}
\nonumber \\
& \qquad \times
  \biggl[ \gamma^\mu\,
   \frac{\slashed{p}' - \bar{u}\slashed{p}
       + \slashed{k}}{\bar{u} Q^2 - k^2}\,
   i\sigma^{\alpha\beta}\ms (a\ms l_u)^{n-1}
\nonumber \\
& \qquad\quad
 + (a\ms l_v)^{n-1}\ms i\sigma^{\alpha\beta}\,
   \frac{\slashed{p} - \bar{v}\slashed{p'}
       + \slashed{k}'}{\bar{v} Q^2 - k'^2}\,
   \gamma^\mu
\phantom{\biggl[ \biggr]}
\nonumber \\
& \qquad\quad
 + i\sigma^{\alpha\beta} a^\mu
   \sum_{j=1}^{n-1} (a\ms l_u)^{j-1}\ms (a\ms l_v)^{n-1-j} \biggr]\,
   a_\alpha b_\beta \, ,
\end{align}
where the last three lines of \eqref{f-initial} respectively correspond to
graphs (a), (b) and (c) in figure~\ref{fig:graphs}.  The factors
\begin{alignat}{4}
  \label{derivative-momenta}
a\ms l_u &= \half\ms (u-\bar{u})\ms ap + \half\ms ap' + a k 
         &&= aP\ms (u - \xi \bar{u}) + a k ,
\nonumber \\[0.2em]
a\ms l_v &= \half\ms (v-\bar{v})\ms ap' +  \half\ms ap + a k'
         &&= aP\ms (v + \xi \bar{v}) + a k'
\nonumber \\[0.2em]
\end{alignat}
come from the derivatives $i\lrpartial a = \half\ms (i\rpartial -
i\lpartial) a$ in the operators \eqref{tensor-no-g} and
\eqref{tensor-one-g}.  The denominator of the gluon propagator in all
three graphs is $\bar{u} \bar{v} Q^2 - (k - k')^2$, and the quark
propagators in graphs (a) and (b) have denominators $\bar{u} Q^2 - k^2$
and $\bar{v} Q^2 - k'^2$.  Note that we are using a Minkowskian scalar
product for the vectors $k$ and $k'$, so that $k^2$, $k'^2$ and $(k-k')^2$
are negative.
In Feynman gauge, the numerator of the gluon propagator is $D_{\lambda\mu}
= g_{\lambda\mu}$ and the fermion trace evaluates to
\begin{align}
  \label{master-trace}
& f(u,v; k, k') = \biggl[ f_1 + f_2
  + \frac{\partial}{\partial k^\alpha}\ms
    \bigl( f_3^\alpha + f_4^\alpha - f_5^\alpha \bigr) \biggr]
\nonumber \\
&\qquad \times (a\ms l_u)^{n-1}\, 
  \frac{\bar{u}}{\bar{u} Q^2 - k^2}\,
  \frac{1}{\bar{u} \bar{v} Q^2 - (k - k')^2}
\nonumber \\[0.2em]
&\quad
 + \biggl[ f_6 - \frac{\partial}{\partial k^\alpha}\ms f_7^\alpha
   \biggr]\;
  \sum_{j=1}^{n-1} (a\ms l_u)^{j-1}\ms (a\ms l_v)^{n-1-j}\;
\nonumber \\
&\qquad \times \frac{1}{Q^2}\,
  \frac{1}{\bar{u} \bar{v} Q^2 - (k - k')^2}
\nonumber \\
&\quad - \biggl\{ u \leftrightarrow v, p \leftrightarrow p',
             k \leftrightarrow k', 
             \frac{\partial}{\partial k} \to 
             \frac{\partial}{\partial k'}\ms , \xi \to -\xi \biggr\}
\nonumber \\[0.1em]
\end{align}
with
\begin{align}
  \label{f-functions}
f_1 &= \Bigl[ (ap) (bp') - (bp) (ap') \Bigr]
       \Bigl( u g(u) - \bar{u}\, v\bar{v}\ms g(v) \Bigr) \, ,
\nonumber \\
f_2 &= \Bigl[ (ap) (bk) - (bp) (ak) \Bigr] u g(u)
\nonumber \\
    &\quad - \Bigl[ (ap') (bk) - (bp') (ak) \Bigr] v\bar{v}\ms g(v) \, ,
\nonumber \\
f_3^\alpha &= \Bigl[ (ap) (bp') - (bp) (ap') \Bigr]\ms  k^\alpha\,
              {u\, v\bar{v}}\, g(v) /{2} \, ,
\nonumber \\
f_4^\alpha &= \Bigl[\ms a^\alpha\ms (bk) - b^\alpha\ms (ak) \Bigr]\ms
              (pp')\, {u\, v\bar{v}}\, g(v) /{2} \, ,
\nonumber \\
f_5^\alpha &= \Bigl[\ms a^\alpha\ms (bp) -  b^\alpha\ms (ap) \Bigr]\ms
              (pp')\, u\bar{u}\, v\bar{v}\ms g(v) \, ,
\nonumber \\
f_6 &= \Bigl[ (ap) (bp') - (bp) (ap') \Bigr]\ms
       (ap)\, (v-\bar{v})\, u\bar{u}\ms g(u) \, ,
\nonumber \\
f_7^\alpha &= \Bigl[\ms a^\alpha\ms (bp) - b^\alpha\ms (ap) \Bigr]\ms
              (pp') (ap')\, u\bar{u}\, v\bar{v}\ms g(v) \, ,
\end{align}
where we have split the result into different terms to facilitate the
subsequent discussion.

In \eqref{master-trace} it is understood that the derivatives
$\partial/\partial k^\alpha$ act also on the vectors $k$ that are implicit
in the functions $f_i^\alpha$ and in the factors $(a\ms l_u)$.  Likewise,
the exchange of variables indicated in the last line of
\eqref{master-trace} applies also to the functions $f_i^{}$, $f_i^\alpha$
and the factors $(a\ms l_u)$ and $(a\ms l_v)$.

One can recognize from the factors $g(u)$ and $g(v)$ in
\eqref{f-functions} that the hard-scattering graphs with the insertion of
the chiral-odd operators \eqref{tensor-no-g} and \eqref{tensor-one-g} pick
out a twist-two distribution amplitude in one of the two pions and a
twist-three distribution amplitude in the other, as anticipated earlier.

%%%%%%%%%%%%%%%%%%%%%%%%%%%%%%%%%%%%%%%%%%%%%%%%%%%%%%%%%%%%%

\section{Extracting the leading terms}
\label{sec:extract}

The factorization formalism is based on an expansion in the small
parameter $\Lambda /Q$, where $\Lambda$ stands for nonperturbative
momentum scales.  In this section we will extract the leading terms in
this expansion.

In the following we will assume that the Gegenbauer series for $g(u)$ in
\eqref{Gegenbauer} converges in the interval $u \in [0,1]$, so that
$\phi(u)$ in \eqref{phi-def} vanishes linearly at the end-points.  The
possibility that this may not hold for low or moderate factorization
scales $\mu$ has been discussed in a number of papers, see for instance
\cite{RuizArriola:2002bp,Radyushkin:2009zg,Polyakov:2009je,Li:2009pr,%
Noguera:2010fe}.
However, the anomalous dimensions $\gamma_n$ in \eqref{evolution} are
positive and increase for $n>0$, and evolution to high scales will
eventually ensure the convergence of \eqref{Gegenbauer} irrespective of
the starting conditions.  Since we are interested in the large-$Q^2$
behavior, the assumption that $g(u)$ is finite at the end-points $u=0$ and
$u=1$ is therefore justified.

Due to the denominators of quark and gluon propagators, the integrals over
$u$ and $v$ in \eqref{mat-el-initial} can be divergent when $k$ and $k'$
are zero.  From \eqref{master-trace} and \eqref{f-functions} we see that
these divergences are at most logarithmic in both $u$ and $v$.  For the
moment we will keep the transverse momenta $k$ and $k'$ fixed, and regard
them as of order $\Lambda \ll Q$ for the purpose of power counting.  We
first identify terms in \eqref{master-trace} that after integration over
$u$ and $v$ vanish like a power of $\Lambda /Q$ (possibly times a power of
$\ln Q/\Lambda$).  We neglect these terms since other contributions will
turn out to be finite or to grow like a power of $\ln Q/\Lambda$ in the
large-$Q^2$ limit.

To simplify expressions, we use that
\begin{equation}
  \label{rot-inv-1}
\int d^2 k\, d^2k'\; k^\alpha\, s(k^2, k'^2, k k') = 0
\end{equation}
and
\begin{align}
  \label{rot-inv-2}
& \int d^2 k\, d^2k'\; k^\alpha k^\beta\, s(k^2, k'^2, k k')
\nonumber \\
& \quad = \frac{1}{2}\, g_T^{\alpha\beta}
    \int d^2 k\, d^2k'\; k^2\, s(k^2, k'^2, k k')
\end{align}
because of rotational invariance in the transverse plane, where $s$ is a
scalar function and
\begin{equation}
g_T^{\alpha\beta} = g^{\alpha\beta}
   - \frac{p^\alpha\ms p'{}^\beta + p'{}^\alpha\ms p^\beta}{pp'} .
\end{equation}
Relations analogous to \eqref{rot-inv-1} and \eqref{rot-inv-2} hold with
one or both of $k^\alpha$, $k^\beta$ replaced by $k'^\alpha$, $k'^\beta$.

We now discuss the different terms of \eqref{master-trace} in turn.  The
reader not interested in the intermediate steps of the argument may skip
forward to eq.~\eqref{master-ff}.  Let us start with the contribution
involving $f_4^\alpha$.  If the derivative $\partial/\partial k$ acts on
the factors $(bk)$ and $(ak)$ in $f_4^\alpha$, the result is proportional
to $\smash{a_\alpha\ms g_T^{\alpha\beta} b_\beta - b_\alpha\ms
  g_T^{\alpha\beta} a_\beta}$ and hence vanishes.  If the derivatives act
on a factor $(ak)$ in $(a\ms l_u)$, one is left with at least two powers
of $k$ or $k'$ in the numerator (a single power giving zero after angular
integration), which are multiplied by a term proportional to
\begin{equation}
  \frac{\bar{u}}{\bar{u} Q^2 - k^2}\,
  \frac{\bar{v}}{\bar{u} \bar{v} Q^2 - (k - k')^2} .
\end{equation}
After integration over $u$ and $v$, this term behaves like $\ln Q/\Lambda$
times an even power of $\Lambda/Q$ and can hence be neglected as well.
The terms where the derivative $\partial/\partial k$ acts on the
propagator denominators are proportional to
\begin{multline}
\frac{(ak) (bk)}{\bar{u} Q^2 - k^2}
     + \frac{(ak) (bk) -(ak') (bk)}{\bar{u} \bar{v} Q^2 - (k - k')^2}
- \bigl\{ a \leftrightarrow b \bigr\}
\\[0.2em]
  = {}- \frac{(ak') (bk) - (bk') (ak)}{\bar{u} \bar{v} Q^2 - (k - k')^2} ,
\end{multline}
which vanishes after angular integration.  The contribution from
$f_4^\alpha$ can hence be neglected altogether.

We proceed with the contributions from $f_5^\alpha$ and $f_7^\alpha$.
When $\partial/\partial k$ acts on a factor $(ak)$ in $(a\ms l_u)$, we
obtain
\begin{multline}
(pp')\ms \Bigl[ (a_\alpha\, g_T^{\alpha\beta} a_\beta) (bp)
           - (a_\alpha\, g_T^{\alpha\beta} b_\beta) (ap) \Bigr]
\\[0.1em]
 = (ap)\ms \Bigl[ (ap) (bp') - (bp) (ap') \Bigr]
\end{multline}
multiplied by an expression that, due to the factors $\bar{u}$ and
$\bar{v}$ in the numerator, gives a finite integral over $u$ and $v$ even
if $k=k'=0$.  If, however, the derivative acts on the propagator
denominators, we obtain a term proportional to
\begin{multline}
  \frac{\bar{u}}{\bar{u} Q^2 - k^2}\,
  \frac{\bar{u} \bar{v}}{\bar{u} \bar{v} Q^2 - (k - k')^2}\,
  \biggl[ \frac{(ak)}{\bar{u} Q^2 - k^2}
\\[0.2em]
  + \frac{(ak) - (ak')}{\bar{u} \bar{v} Q^2 - (k - k')^2} \biggr] (bp)
- \bigl\{ a \leftrightarrow b \bigr\} .
\end{multline}
At least one more power of $(ka)$ from $(a\ms l_u)$ is required to get a
nonvanishing term after angular integration.  The integrals over $u$ and
$v$ are only logarithmically divergent, so that this contribution is
suppressed by an even power of $\Lambda/Q$ and can again be neglected.

Let us now discuss the term with $f_3^\alpha$.  The contribution from the
derivative $\partial/\partial k$ acting on $k^\alpha$ needs to be
retained, whereas contributions with the derivative acting on a factor
$(ak)$ in $(a\ms l_u)$ can be neglected: they have at least two powers of
$k$ in the numerator, which are multiplied by an expression that gives
only a logarithm $\ln Q/\Lambda$ after integration over $u$ and $v$.  When
the derivative acts on the propagator denominators, we get a term
proportional to
\begin{multline}
  \label{first-line}
  \frac{\bar{u}}{\bar{u} Q^2 - k^2}\,
  \frac{\bar{v}}{\bar{u} \bar{v} Q^2 - (k - k')^2}\,
\biggl[ \frac{k^2}{\bar{u} Q^2 - k^2}
\\[0.2em]
      + \frac{k\ms (k-k')}{\bar{u} \bar{v} Q^2 - (k - k')^2} \biggr] .
\end{multline}
The integral over $v$ of this term gives a logarithm $\ln Q/\Lambda$,
whereas the one over $u$ diverges linearly for $k=k'=0$.  For finite $k$
and $k'$ the $u$-integral thus provides a factor $1/\Lambda^2$ that
cancels the factor $\Lambda^2$ from the transverse momenta in the
numerator.  Note, however, that the expression in \eqref{first-line} is
multiplied by $n-1$ powers of $(a\ms l_u) = aP\ms (u - \xi \bar{u}) + a
k$.  Only the contributions from $(aP) u$ need to be retained, since a
factor $\bar{u}$ turns the linearly divergent $u$-integral of
\eqref{first-line} into a logarithmically divergent one, whereas factors
of $(ak)$ directly provide further powers of $(\Lambda/Q)^2$.

After performing the derivatives $\partial/\partial k$ and
$\partial/\partial k'$ in \eqref{master-trace}, we can omit all terms
$(ak)$ in $(a\ms l_u)$ and $(ak')$ in $(a\ms l_v)$, since they give rise
to power suppressed terms.  Furthermore, the contribution from $f_2$ is
power suppressed and can be neglected.

Putting everything together we have
\begin{align}
  \label{second-trace}
& \int du\, dv\, \;d^2 k\, d^2k'\; \Sigma(u,k^2)\,
  \Sigma(v,k'^2)\, f(u,v; k, k')
\nonumber \\
&\ = \Bigl[ (ap) (bp') - (bp) (ap') \Bigr]\ms (aP)^{n-1}
     \phantom{\int}
\nonumber \\
&\quad \times
  \int du\, dv\; d^2 k\, d^2k'\; \Sigma(u,k^2)\, \Sigma(v,k'^2)\;
\nonumber \\
&\quad \times
   \frac{1}{Q^2}\, \frac{1}{\bar{u} \bar{v} Q^2 - (k - k')^2} \,
   \biggl( \frac{\bar{u} Q^2}{\bar{u} Q^2 - k^2}
\nonumber \\
&\qquad \times \biggl\{
   \Bigl[ u g(u) + (u - \bar{u})\, v\bar{v}\ms g(v) \Bigr] \ms
   (u - \xi\bar{u})^{n-1}
\nonumber \\
&\qquad\quad 
  + u^{n}\ms v\bar{v}\ms g(v)\, \biggl[ \frac{k^2}{\bar{u} Q^2 - k^2}
           + \frac{k\ms (k-k')}{\bar{u} \bar{v} Q^2 - (k - k')^2} \biggr]
\nonumber \\
&\qquad\quad - (1+\xi)\, u\bar{u}\, v\bar{v}\ms g(v)\ms
                 (n-1) (u - \xi\bar{u} )^{n-2} \ms\biggr\} \;
\nonumber \\
&\qquad + (1+\xi) (v-\bar{v})\, u\bar{u}\ms g(u)
        \phantom{\frac{1}{1}}
\nonumber \\
&\qquad\quad \times 
     \sum_{j=1}^{n-1} (u - \xi\bar{u})^{j-1}\ms (v + \xi\bar{v})^{n-1-j}
\nonumber \\ 
&\qquad - (1-\xi^2)\, u\bar{u}\, v\bar{v}\ms g(v)
        \phantom{\frac{1}{1}}
\nonumber \\
&\qquad\quad \times \sum_{j=1}^{n-1} (j-1) (u - \xi\bar{u})^{j-2}\ms
     (v + \xi\bar{v})^{n-1-j} \ms\biggr)
\nonumber \\
&\quad + \Bigl\{ u \leftrightarrow v,
             k \leftrightarrow k', \xi \to -\xi \Bigr\}
  + \mathcal{O}\biggl( \frac{\Lambda^2}{Q^2}\,
                       \ln^2 \frac{Q^2}{\Lambda^2} \biggr) .
\nonumber \\[0.1em]
\end{align}
Before proceeding let us mention that we checked the gauge independence of
our result for a general covariant gauge.  Using the same methods as those
leading to \eqref{second-trace}, we find that the gauge dependent part of
$D_{\lambda\mu}$ gives only contributions suppressed by an even power of
$\Lambda/Q$.

Let us now rewrite \eqref{second-trace} in a form that allows us to
identify those terms that give logarithms in $Q/\Lambda$.  For the term
proportional to $v\bar{v}\ms g(v)$ in the fifth line of
\eqref{second-trace} we can write
\begin{align}
 & (u - \bar{u})\, (u - \xi\bar{u})^{n-1} \phantom{\Bigl[ \Bigr]}
\nonumber \\
 &\quad = 1 - 2\bar{u} (u - \xi\bar{u})^{n-1}
    - \Bigl[ 1 - (u - \xi\bar{u})^{n-1} \Bigr]
\nonumber \\
 &\quad = 1 - 2\bar{u} (u - \xi\bar{u})^{n-1}
      - (1+\xi)\, \bar{u} \sum_{j=1}^{n-1} (u - \xi\bar{u})^{j-1} ,
\end{align}
where in the last step we have used the geometric series.  Similarly, the
terms proportional to $u g(u)$ in \eqref{second-trace} can be rewritten as
\begin{align}
& (u - \xi\bar{u})^{n-1}\, \frac{\bar{u} Q^2}{\bar{u} Q^2 - k^2}
  \phantom{\sum_{j}^{1}}
\nonumber \\
& \quad + (1+\xi)\, (v-\bar{v})\, \bar{u}
  \sum_{j=1}^{n-1} (u - \xi\bar{u})^{j-1}\ms (v + \xi\bar{v})^{n-1-j}
\nonumber \\  
& = (u - \xi\bar{u})^{n-1}\, \frac{\bar{u} Q^2}{\bar{u} Q^2 - k^2}
  + (1+\xi)\, \bar{u} \sum_{j=1}^{n-1} (u - \xi\bar{u})^{j-1}
\nonumber \\  
&\quad {}- (1+\xi)\, \bar{u} \sum_{j=1}^{n-1} (u - \xi\bar{u})^{j-1}
         \Bigl[ 1 - (v + \xi\bar{v})^{n-1-j} \Bigr]
\nonumber \\
& \quad {}- 2 (1+\xi)\, \bar{v} \bar{u}
    \sum_{j=1}^{n-1} (u - \xi\bar{u})^{j-1}\ms (v + \xi\bar{v})^{n-1-j}
\nonumber \\  
&= 1 + (u - \xi\bar{u})^{n-1}\, \frac{k^2}{\bar{u} Q^2 - k^2}
  \phantom{\sum_{j}^{1}}
\nonumber \\
& \quad {}- (1-\xi^2)\, \bar{u} \bar{v}
     \sum_{j=1}^{n-1} (u - \xi\bar{u})^{j-1}
     \sum_{l=1}^{n-1-j} (v + \xi\bar{v})^{l-1}
\nonumber \\  
&\quad {}- 2 (1+\xi)\, \bar{u}\bar{v}
    \sum_{j=1}^{n-1} (u - \xi\bar{u})^{j-1}\ms (v + \xi\bar{v})^{n-1-j} .
\end{align}
In the term proportional to $k^2$ we only need to keep the factor
$u^{n-1}$, since with one or more factors of $\xi\bar{u}$ we get only a
logarithmically divergent integral over $u$ and $v$ multiplied by $k^2$,
which is power suppressed.  Finally, we observe that for those terms in
the large braces of \eqref{second-trace} that contain a factor
$\bar{u}\bar{v}$, we have
\begin{equation}
  \label{prop-approx}
\frac{\bar{u}\bar{v} Q^2}{\bar{u} \bar{v} Q^2 - (k - k')^2}\,
\frac{\bar{u} Q^2}{\bar{u} Q^2 - k^2}
  = 1 + \mathcal{O}\biggl( \frac{\Lambda^2}{Q^2} \biggr) \, .
\end{equation}
Using the definition \eqref{ff-def} of the form factors we then obtain
\begin{align}
  \label{master-ff}
 & \sum_{\genfrac{}{}{0pt}{}{i=0}{\text{even}}}^{n-1}
    (2\xi)^i\ms B_{T ni}^{u}(Q^2) =
4\pi\alpha_s \frac{C_F}{N_c}\, 
   \frac{6 f_\pi^2\ms m_\pi^{} \mu_\pi^{}}{Q^4}
\nonumber \\
 &\quad \times  \int du\, dv\; d^2 k\, d^2k'\;
   \Sigma(u,k^2)\, \Sigma(v,k'^2)
\nonumber \\
&\quad \times \biggl( \frac{Q^2}{\bar{u} \bar{v} Q^2 - (k - k')^2}\;
  \biggl\{ u g(u) 
         + v\bar{v}\ms g(v)\ms \frac{\bar{u} Q^2}{\bar{u} Q^2 - k^2}
\nonumber \\ 
&\hspace{4.2em}
  + u^{n} g(u)\, \frac{k^2}{\bar{u} Q^2 - k^2}
  + u^{n}\ms v\bar{v}\ms g(v)\, 
    \frac{\bar{u} Q^2}{\bar{u} Q^2 - k^2}
\nonumber \\
&\hspace{7em} \times 
    \biggl[ \frac{k^2}{\bar{u} Q^2 - k^2}
       + \frac{k\ms (k-k')}{\bar{u} \bar{v} Q^2 - (k - k')^2} \biggr]
 \biggr\}
\nonumber \\[0.5em]
&\qquad - 2 v g(v)\ms (u - \xi\bar{u})^{n-1}
\nonumber \\
&\qquad - (1+\xi)\, v g(v) \sum_{j=1}^{n-1} (u - \xi\bar{u})^{j-1}
\nonumber \\
&\qquad - (1+\xi)\, v g(v)\ms (n-1)\ms u (u - \xi\bar{u} )^{n-2}
\nonumber \\
&\qquad - (1-\xi^2)\, u g(u)
     \sum_{j=1}^{n-1} (u - \xi\bar{u})^{j-1} \!
     \sum_{l=1}^{n-1-j} (v + \xi\bar{v})^{l-1}
\nonumber \\
&\qquad - 2 (1+\xi)\, u g(u)
   \sum_{j=1}^{n-1} (u - \xi\bar{u})^{j-1}\ms (v + \xi\bar{v})^{n-1-j}
\nonumber \\[0.2em]
&\qquad - (1-\xi^2)\, v g(v)
\nonumber \\
&\qquad\quad \times
   \sum_{j=1}^{n-1} (j-1)\ms u (u - \xi\bar{u})^{j-2}\ms
     (v + \xi\bar{v})^{n-1-j} \ms\biggr)
\nonumber \\
&\quad
  + \Bigl\{ u \leftrightarrow v,
             k \leftrightarrow k', \xi \to -\xi \Bigr\}
  + \mathcal{O}\biggl( \frac{\Lambda^2}{Q^2}\,
                       \ln^2 \frac{Q^2}{\Lambda^2} \biggr)
\nonumber \\
&= 4\pi\alpha_s \frac{C_F}{N_c}\, 
   \frac{6 f_\pi^2\ms m_\pi^{} \mu_\pi^{}}{Q^4}\,
   \int du\, dv\; d^2 k\, d^2k'\;
     \Sigma(u,k^2)\,
\nonumber \\
&\quad \times \Sigma(v,k'^2)\;
  \frac{2 Q^2}{\bar{u} \bar{v} Q^2 - (k - k')^2}\;
  \biggl\{ g(u) \bigl( 1 - \bar{u}^2 \bigr)
\nonumber \\ 
&\qquad
  + \Bigl[ v\bar{v}\ms g(v)
         + u^{n}\ms g(u) \Bigr] \frac{k^2}{\bar{u} Q^2 - k^2}
  + u^{n}\ms v\bar{v}\ms g(v)\,
\nonumber \\
&\qquad\quad \times  \frac{\bar{u} Q^2}{\bar{u} Q^2 - k^2}\,
 \biggl[ \frac{k^2}{\bar{u} Q^2 - k^2}
       + \frac{k\ms (k-k')}{\bar{u} \bar{v} Q^2 - (k - k')^2} \biggr]
 \biggr\}
\nonumber \\[0.5em]
&\quad - 4\pi\alpha_s \frac{C_F}{N_c}\, 
   \frac{6 f_\pi^2\ms m_\pi^{} \mu_\pi^{}}{Q^4}\,
   \int du\, dv\; u g(u)
\nonumber \\
&\qquad \times \biggl(
    2 (v + \xi\bar{v})^{n-1}
    + (1-\xi)\ms \biggl[ (n-1)\ms v (v + \xi\bar{v})^{n-2}
\nonumber \\
&\hspace{15.5em}
    + \sum_{j=1}^{n-1} (v + \xi\bar{v})^{j-1} \biggr]
\nonumber \\
&\qquad\quad + (1+\xi)\, \sum_{j=1}^{n-1} (u - \xi\bar{u})^{n-1-j}\ms
  \biggl\{\ms 2 (v + \xi\bar{v})^{j-1}
\nonumber \\
&\qquad\qquad
   + (1-\xi)\ms \biggl[ (j-1)\ms v (v + \xi\bar{v})^{j-2}
\nonumber \\
&\hspace{10em}
   + \sum_{l=1}^{j-1} (v + \xi\bar{v})^{l-1} \biggr] \biggr\}
\nonumber \\
&\qquad\quad
  + \Bigl\{ \xi \to -\xi \Bigr\}\ms \biggr)
  + \mathcal{O}\biggl( \frac{\Lambda^2}{Q^2}\,
                       \ln^2 \frac{Q^2}{\Lambda^2} \biggr) ,
\end{align}
where in the last step we have changed the summation index $j\to n-j$ in
the double sum.  For the terms where the quark and gluon propagators have
canceled, we performed the integrations over $k$ and $k'$ using the
normalization condition \eqref{sigma-norm} for $\Sigma$.

{}From \eqref{master-ff} we read off an important result:
\begin{enumerate}
\item The $\xi$ dependent terms of the matrix element \eqref{ff-def} and
  thus the form factors $B_{T ni}^{u}$ with $i\ge 2$ behave like $1/Q^4$
  at large $Q$, up to logarithmic corrections from the dependence of
  $\alpha_s$, $\mu_\pi$ and $g(u)$ on the renormalization or factorization
  scale, which one should take proportional to $Q^2$.

  These form factors can be calculated in standard collinear
  factorization, and the regulating functions $\Sigma(u, k^2)\, \Sigma(v,
  k'^2)$ we used in the intermediate steps of our calculation have
  completely disappeared.  The reason for this can be traced back to
  \eqref{master-trace}, where the only $\xi$ dependence comes from the
  factors $(a\ms l_u)$ and $(a\ms l_v)$ and is accompanied by factors
  $\bar{u}$ or $\bar{v}$ according to \eqref{derivative-momenta}.  These
  factors suppress the end-point regions and turn out to make the $u$ and
  $v$ integrals finite in the collinear limit $k=k'=0$.
\item The form factors $B_{T n0}^{u}$ involve logarithmically divergent
  integrals over $u$ and $v$ in the collinear limit and thus give rise to
  logarithms of $Q/\Lambda$ if we regularize these divergences.
\end{enumerate}
In the following subsections we shall discuss the two cases in turn.

Before doing so, let us comment on the behavior of our result
\eqref{master-ff} in the limit of vanishing pion mass.  The parameter
$\mu_\pi$, which originates from the pion projection operator
\eqref{projectors}, is proportional to the chiral condensate and remains
finite in the chiral limit.  According to \eqref{master-ff} the form
factors $B_{T ni}^u$ therefore vahish like $m_\pi$ in that limit, which is
simply due to the factor $1/m_\pi$ multiplying them in their definition
\eqref{ff-def}.  The pion matrix element in \eqref{ff-def} remains finite
in the chiral limit.  Note finally that when calculating the hard
scattering we have neglected the quark masses, which are small not only
compared with $Q$ but also compared with the typical values of transverse
quark momenta, which we have retained in the denominators of propagators
to avoid divergent integrals.

%%%%%%%%%%%%%%%%%%%%%%%%%%%%%%%%%%

\subsection{The form factors $B_{T ni}^{u}$ with $i\ge 2$}

{}From \eqref{master-ff} one can readily extract the expressions for the
form factors $B_{T ni}^{u}$ with $i\ge 2$.  The integrals over $v$ are
elementary, as well as those over $u$ if $g(u)$ is explicitly given as a
Gegenbauer series \eqref{Gegenbauer}.  For general $n$ and $k$ the
expressions become rather lengthy, but they remain short for the term
$k=n-1$ with the maximal power of $\xi$.  We obtain
\begin{align}
  \label{Bnn1}
& B_{T n,n-1}^{u} = 4 \pi\alpha_s \frac{C_F}{N_c}\, 
   \frac{6 f_\pi^2\ms m_\pi^{} \mu_\pi^{}}{Q^4}\, \frac{1}{2^{n-2}}
   \int du\, dv\; u g(u)\,
\nonumber \\
&\quad \times
   \biggl\{ n \bar{v}^{n-2} - (n+1) \bar{v}^{n-1} + 2 \bar{u}^{\ms n-2}
\nonumber \\   
&\qquad\ + \sum_{j=2}^{n-1} (-\bar{u})^{n-1-j}\ms
   \Bigl[\ms j\ms \bar{v}^{\ms j-2} - (j+1)\ms \bar{v}^{\ms j-1} \Bigr]
   \biggr\}
\nonumber \\
&\ = 4 \pi\alpha_s \frac{C_F}{N_c}\, 
   \frac{6 f_\pi^2\ms m_\pi^{} \mu_\pi^{}}{Q^4}\, \frac{1}{2^{n-2}}
   \int du\, u g(u)\,
\nonumber \\
&\quad \ \times
   \biggl\{ \frac{1}{n(n-1)} + 2 \bar{u}^{\ms n-2}
         + \sum_{j=2}^{n-1} \frac{(-\bar{u})^{n-1-j}}{j(j-1)} \ms\biggr\} ,
\end{align}
where $n \ge 3$ must be odd.  For $n=3$ this gives
\begin{align}
  \label{B32}
B_{T 32}^{u} &= 4 \pi\alpha_s \frac{C_F}{N_c}\, 
    \frac{6 f_\pi^2\ms m_\pi^{} \mu_\pi^{}}{Q^4}\,
    \int du\, u g(u)\ms \biggl( \frac{1}{3} + \bar{u} \biggr)
\nonumber \\
&= 4 \pi\alpha_s \frac{C_F}{N_c}\, 
    \frac{f_\pi^2\ms m_\pi^{} \mu_\pi^{}}{Q^4}\,
  \biggl( 2 + \sum_{n=2}^\infty a_n \biggr) .
\end{align}
These expressions hold up to power corrections in $\Lambda^2 /Q^2$ and to
leading order in $\alpha_s$.

%%%%%%%%%%%%%%%%%%%%%%%%%%%%%%%%%%

\subsection{The form factors $B_{T n0}^{u}$}

The form factors $B_{T n0}^{u}$ correspond to the $\xi$-independent part
of \eqref{master-ff}.  Let us first take a closer look at terms that have
a factor $k^2$ or $k (k-k')$ in the numerator.  By explicit integration we
find that the integrals $\int du\, dv$ of
\begin{align}
& \frac{1}{\bar{u} \bar{v} Q^2 - (k - k')^2}\,
  \frac{k^2}{\bar{u} Q^2 - k^2} \,,
\nonumber \\
& \frac{1}{\bar{u} \bar{v} Q^2 - (k - k')^2}\,
    \frac{k^2}{\bar{u} Q^2 - k^2}\,
    \frac{\bar{u}}{\bar{u} Q^2 - k^2} \,,
\nonumber \\
& \frac{1}{\bar{u} \bar{v} Q^2 - (k - k')^2}\,
  \frac{k(k-k')}{\bar{u} \bar{v} Q^2 - (k - k')^2}\,
  \frac{\bar{u}}{\bar{u} Q^2 - k^2}
\end{align}
are finite for $k=k'=0$, as well as the corresponding integrals with extra
factors of $\bar{u}$ and $\bar{v}$ in the numerator.  We thus have
\begin{align}
  \label{ff-kperp}
& B_{T n0}^{u}(Q^2) =
8\pi\alpha_s \frac{C_F}{N_c}\, 
   \frac{6 f_\pi^2\ms m_\pi^{} \mu_\pi^{}}{Q^4}\;
  \biggl\{ \int du\, dv\; d^2 k\, d^2k'
\nonumber \\[0.1em]
& \quad \times \Sigma(u,k^2)\, \Sigma(v,k'^2)\, 
   \frac{g(u) ( 1 - \bar{u}^2 )}{\bar{u} \bar{v} +
                                 (\tvec{k}-\tvec{k}')^2 /Q^2}
 + \mathcal{O}(1) \biggr\} ,
\nonumber \\[0.2em]
\end{align}
where the boldface symbols indicate that we are now using a Euclidean
scalar product in transverse momentum space, i.e.\ $(k-k')^2 = -
(\tvec{k}-\tvec{k}')^2$.  Remarkably, the r.h.s.\ of \eqref{ff-kperp} is
independent of $n$, i.e.\ the contribution enhanced by powers of $\ln
Q/\Lambda$ is the \emph{same} for all $n$.  The contribution indicated as
$\mathcal{O}(1)$ does not develop logarithms of $Q/\Lambda$ and depends on
$n$, as is obvious from \eqref{master-ff}.

To proceed, we replace $(\tvec{k}-\tvec{k}')^2$ in \eqref{ff-kperp} by a
constant $\Lambda^2$, which thus plays the role of a typical squared
transverse momentum in the gluon propagator.  With the normalization
condition \eqref{sigma-norm} for $\Sigma$ this replacement gives
\begin{align}
  \label{k-to-lambda}
\int d^2 k\, d^2k'\; 
  \frac{\Sigma(u,k^2)\, \Sigma(v,k'^2)}{\bar{u} \bar{v} +
                                        (\tvec{k}-\tvec{k}')^2 /Q^2}
&\to \frac{1}{\bar{u} \bar{v} + \Lambda^2 /Q^2} \, .
\end{align}
Clearly, this is an oversimplification since in general the average value
of $(\tvec{k}-\tvec{k}')^2$ in the integral will depend on $u$ and $v$ and
cannot be described by a single constant $\Lambda^2$.  However, we
consider \eqref{k-to-lambda} as sufficient for our purpose, bearing also
in mind that even the description of the transverse-momentum dependence by
a single function $\Sigma(u,k^2)$ is a simplified ansatz, as discussed
after eq.~\eqref{sigma-norm}.

After the replacement \eqref{k-to-lambda} we can perform the $v$
integration in \eqref{ff-kperp} and get
\begin{align}
  \label{ff-lambda}
 B_{T n0}^{u}(Q^2) &=
   8\pi\alpha_s \frac{C_F}{N_c}\, 
   \frac{6 f_\pi^2\ms m_\pi^{} \mu_\pi^{}}{Q^4}\,
  \biggl\{ \int_0^1 du\, g(u)
\nonumber \\
& \times \bigl( 1 - \bar{u}^2)\, \frac{1}{\bar{u}} 
         \ln\frac{\bar{u} Q^2 + \Lambda^2}{\Lambda^2}
 + \mathcal{O}(1) \biggr\} \, .
\end{align}
To make the logarithms of $Q/\Lambda$ explicit we use that
\begin{align}
  \label{dilog}
& \int_0^1 du\, \frac{1}{\bar{u}}
              \ln\frac{\bar{u} Q^2 + \Lambda^2}{\Lambda^2}
\nonumber \\
& \quad
  = - \operatorname{Li}_2 \Bigl(- \frac{Q^2}{\Lambda^2} \Bigr)
  = \frac{1}{2} \ln^2\frac{Q^2}{\Lambda^2} + \mathcal{O}(1)
\intertext{and}
& \int_0^1 du\,
   r(\bar{u})\ms \ln\frac{\bar{u} Q^2 + \Lambda^2}{\Lambda^2}
\nonumber \\
&\quad = \int_0^1 du\,  r(\bar{u})\, \biggl[ \ln\frac{Q^2}{\Lambda^2}
     + \ln\Bigl( \bar{u} + \frac{\Lambda^2}{Q^2} \Bigr) \biggr]
\nonumber \\
&\quad = \ln\frac{Q^2}{\Lambda^2} \int_0^1 du\, r(\bar{u})
     + \mathcal{O}(1)
\end{align}
if $r(\bar{u})$ is finite at $\bar{u}=0$.  We note that the term of
$\mathcal{O}(1)$ in \eqref{dilog} is equal to $\pi^2 /6 \approx 3.3 /2$,
so that one should only use our approximation for $\ln^2 (Q^2/\Lambda^2)
\gg 3.3$.
Our final result then reads
\begin{align}
  \label{master-result}
& B_{T n0}^{u} =
   4\pi\alpha_s \frac{C_F}{N_c}\, 
   \frac{6 f_\pi^2\ms m_\pi^{} \mu_\pi^{}}{Q^4}\;
\biggl\{ g(1)\ms \ln^2\frac{Q^2}{\Lambda^2}
         - 2\ms \ln\frac{Q^2}{\Lambda^2}
\nonumber \\[0.1em]
& \qquad\qquad \times \int_0^1 du\, 
         \biggl[ \frac{g(u) - g(1)}{u-1} + \bar{u}\ms g(u) \biggr] 
  + \mathcal{O}(1) \biggr\}
\nonumber \\[0.5em]
&\quad = 24\pi\alpha_s \frac{C_F}{N_c}\, 
   \frac{f_\pi^2\ms m_\pi^{} \mu_\pi^{}}{Q^4}\;
\nonumber \\
&\qquad \times \biggl\{ \ln^2\frac{Q^2}{\Lambda^2}\,
   \bigl( 1 + 6\ms a_2 + 15\ms a_4 + 28\ms a_6 + \cdots \bigr)
\nonumber \\
&\qquad\quad  - \ln\frac{Q^2}{\Lambda^2}\,
   \bigl( 1 + 31\ms a_2 + 106\ms a_4 + 233.4\ms a_6 + \cdots \bigr)
\nonumber \\
&\qquad\quad  + \mathcal{O}(1) \biggr\} ,
\end{align}
where $\bigl[g(u) - g(1)\bigr] \big/ (u-1)$ is finite at $u=1$.

In stark contrast to the case of $B_{T n,n-1}^u$ in \eqref{Bnn1} and
\eqref{B32}, the result \eqref{master-result} depends very strongly on the
end-point behavior of the twist-two pion distribution amplitude $\phi(u)$,
or in other words on the higher Gegenbauer coefficients $a_n$ in the
expansion \eqref{Gegenbauer}.  One can expect that Sudakov effects will
weaken this dependence by suppressing the end-points in $u$, but to
investigate this is beyond the scope of the present work.  One should,
however, be wary to take the strong end-point dependence in
\eqref{master-result} at face value.

%%%%%%%%%%%%%%%%%%%%%%%%%%%%%%%%%%%%%%%%%%%%%%%%%%%%%%%%%%%%%

\section{Numerical illustration}
\label{sec:numerical}

In this section we give some numerical illustrations of our results.  This
is to obtain a basic feeling for the order of magnitude and the $Q^2$
behavior of our expressions \eqref{B32} and \eqref{master-result}.  To
provide a baseline, we also plot the electromagnetic pion form factor,
calculated in the same approximation as \eqref{B32}, i.e.\ in collinear
factorization at leading order in $\alpha_s$:
\begin{align}
  \label{fpi}
F_\pi(Q^2) &= 18\pi\alpha_s \frac{C_F}{N_c}\, 
              \frac{f_\pi^2}{Q^2}\, \biggl[ \int du\, g(u) \ms\biggr]^2
\nonumber \\
 &= 18\pi\alpha_s \frac{C_F}{N_c}\, 
    \frac{f_\pi^2}{Q^2}\, \biggl( 1 + \sum_{n=2}^\infty a_n \biggr)^2 .
\end{align}
At experimentally relevant values of $Q^2$ the result \eqref{fpi} receives
important corrections from higher orders in $\alpha_s$ and from various
types of power corrections
\cite{Nesterenko:1982gc,Li:1992nu,Jakob:1993iw,Braun:1999uj,%
  Bakulev:2004cu,Melic:1998qr}.  It is natural to expect the same of our
result for $B_{T 32}^u$, and even more so for $B_{T n0}^u$, where the
strictly collinear framework is not applicable.

In the following we use the one-loop expression for $\alpha_s$ with
$n_F=4$ active quark flavors and $\smash{\Lambda_{QCD}^{(4)}} = 181 \mev$.
This gives $\alpha_s(m_\tau) = 0.33$ in agreement with extractions of the
strong coupling form $\tau$ decays \cite{Bethke:2009jm}.  For the quark
masses we take the value $(m_u + m_d)/2 = 3.79 \mev$ at the scale $\mu_0=
2\gev$ \cite{Amsler:2008zzb}, which according to \eqref{chiral-param}
results in $\mu_\pi = 2.57 \gev$ at the same scale.  To illustrate the
dependence on the twist-two distribution amplitude, we take either its
asymptotic form $\phi(u) = 6 u\bar{u}$ or a form with $a_2 = 0.2$ at
$\mu_0= 2 \gev$ and all other Gegenbauer coefficients set to zero.  The
value of $a_2$ just quoted is close to what has been obtained in two
recent lattice calculations \cite{Braun:2006dg,Donnellan:2007xr}.  The
one-loop scale dependence of $\mu_\pi$ and $a_n$ is given in
\eqref{evolution}, in particular one finds that $\mu_\pi(\mu)$ behaves
like $\alpha_s(\mu)^{-0.48}$ for $n_F=4$.

\begin{figure}[th!]
\begin{center}
\includegraphics[width=0.49\textwidth]{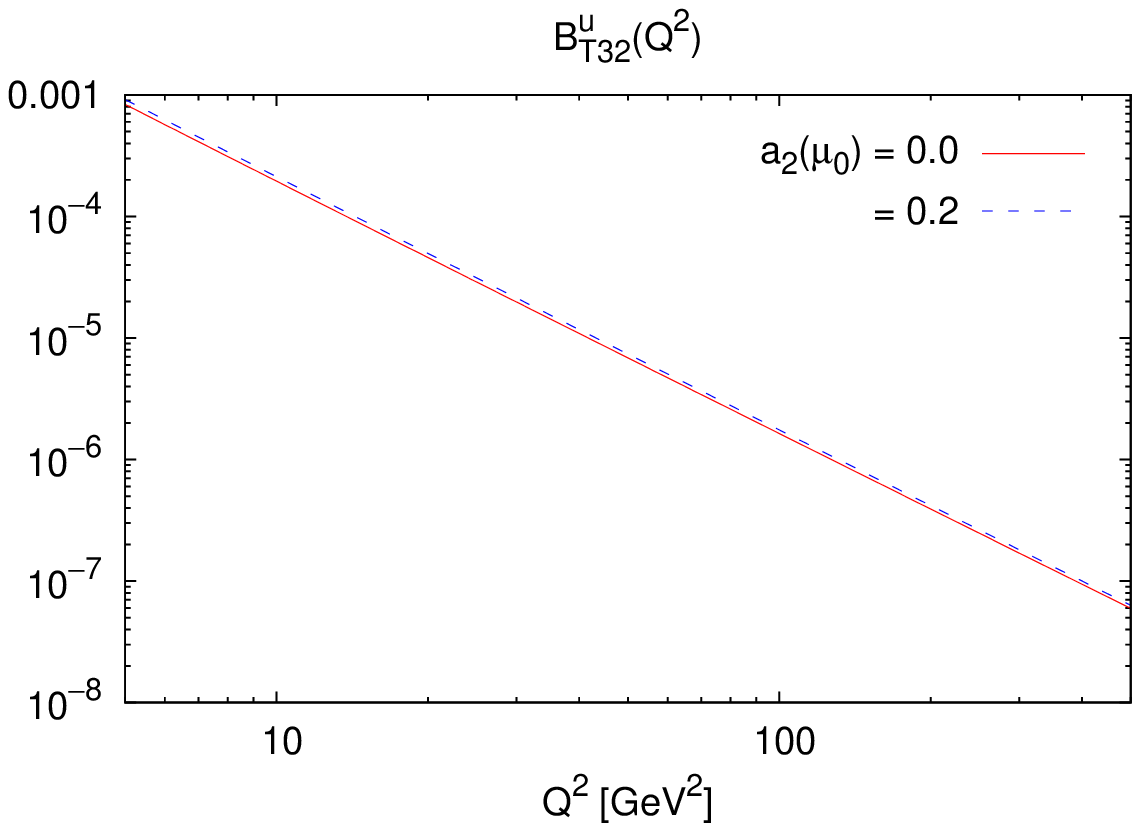} \\[1em]
\includegraphics[width=0.49\textwidth]{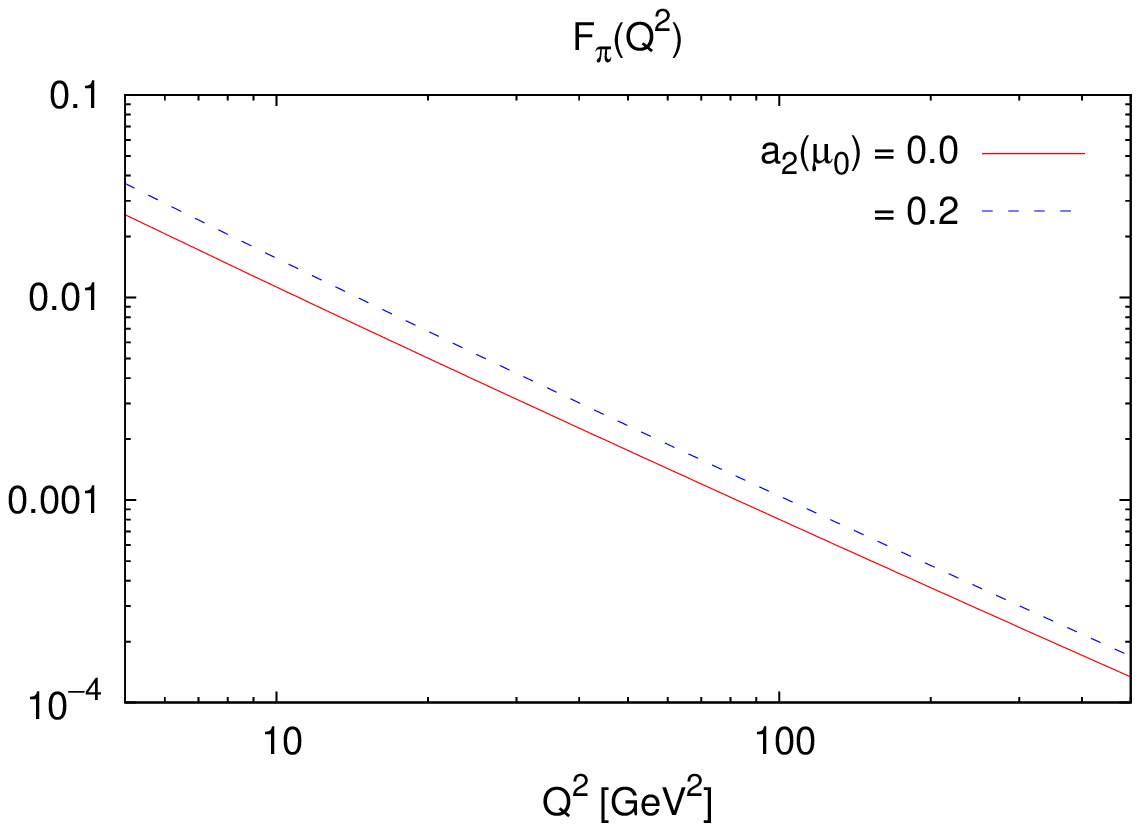}
\end{center}
\caption{\label{fig:B32} The form factors $B_{T 32}^u$ and $F_\pi$ in
  collinear factorization, as given in \protect\eqref{B32} and
  \protect\eqref{fpi}.  The factorization and renormalization scales are
  set to $\mu=Q$.  The solid (dashed) curve is for $a_2 = 0$ ($0.2$) at
  $\mu_0=2\gev$, with all other Gegenbauer coefficients set to zero.}
\end{figure}

In Fig.~\ref{fig:B32} we show our result \eqref{B32} for $B_{T 32}^u$
along with $F_\pi$.  We have taken $\mu^2 = Q^2$ for the renormalization
and factorization scales.  For a baseline estimate this is a natural
choice, and we will not explore here the more sophisticated options
discussed in the literature \cite{Bakulev:2004cu,Melic:1998qr}.
We see in the figure that $B_{T 32}^u$ is over an order of magnitude
smaller than $F_\pi$ already at $Q^2 = 5\gev^2$.  Of course, the
difference between these form factors increases with $Q^2$ because of
their different power behavior.  We note that both $B_{T 32}^2(Q^2)$ and
$F_\pi(Q^2)$ decrease slightly faster than their nominal powers $1/Q^4$
and $1/Q^2$.  This is due to the running of $\alpha_s$, which in the case
of $B_{T 32}^u$ is more important than the increase of $\mu_\pi$ with the
factorization scale.
We finally observe that the dependence on the Gegenbauer coefficient $a_2$
is weaker for $B_{T 32}$ than for $F_\pi$, which is readily understood
from the respective expressions \eqref{B32} and \eqref{fpi}.

\begin{figure}
\begin{center}
\includegraphics[width=0.49\textwidth]{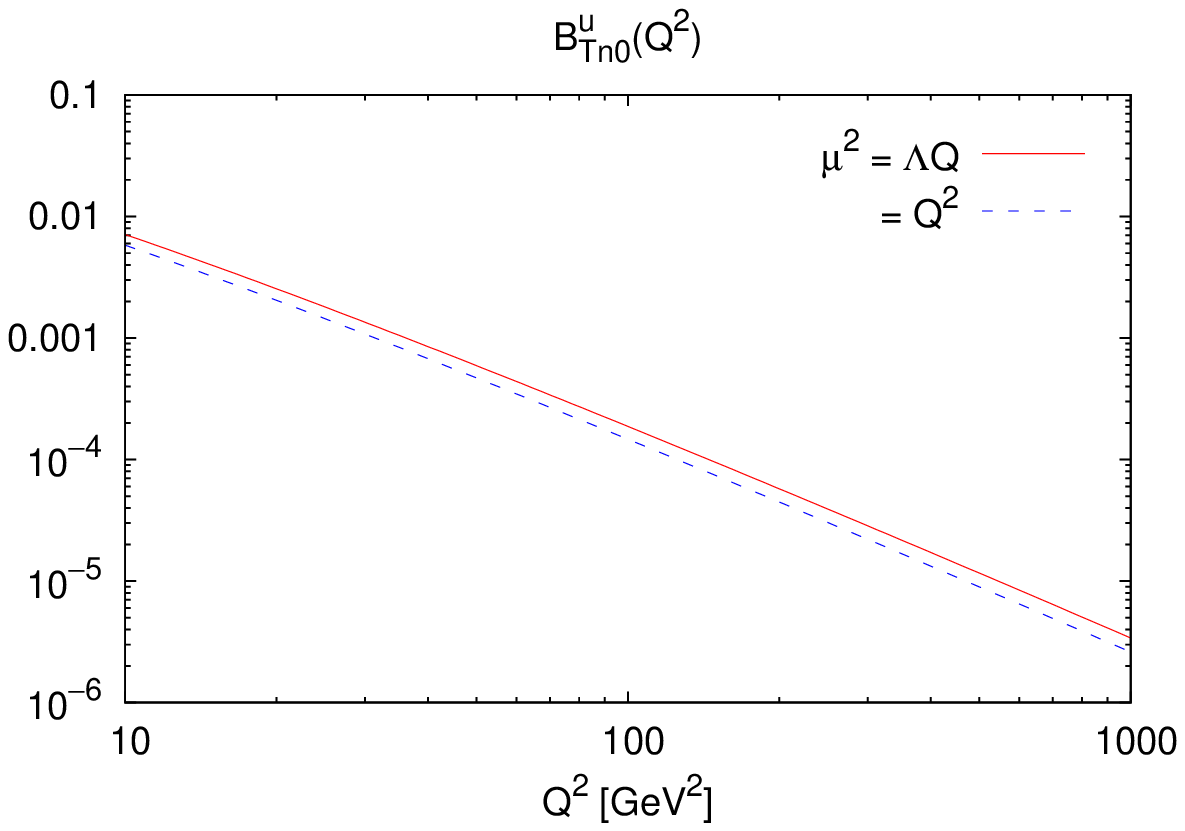} \\[1em]
\includegraphics[width=0.49\textwidth]{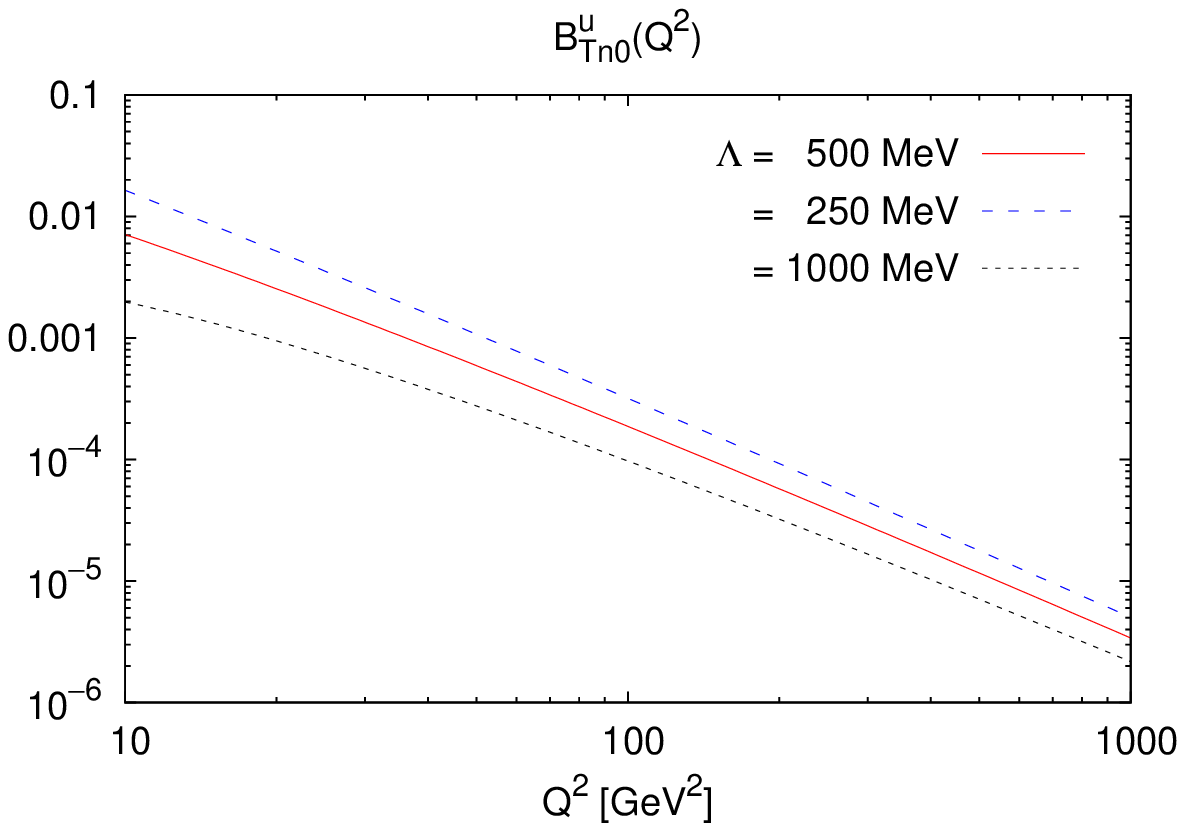} \\[1em]
\includegraphics[width=0.49\textwidth]{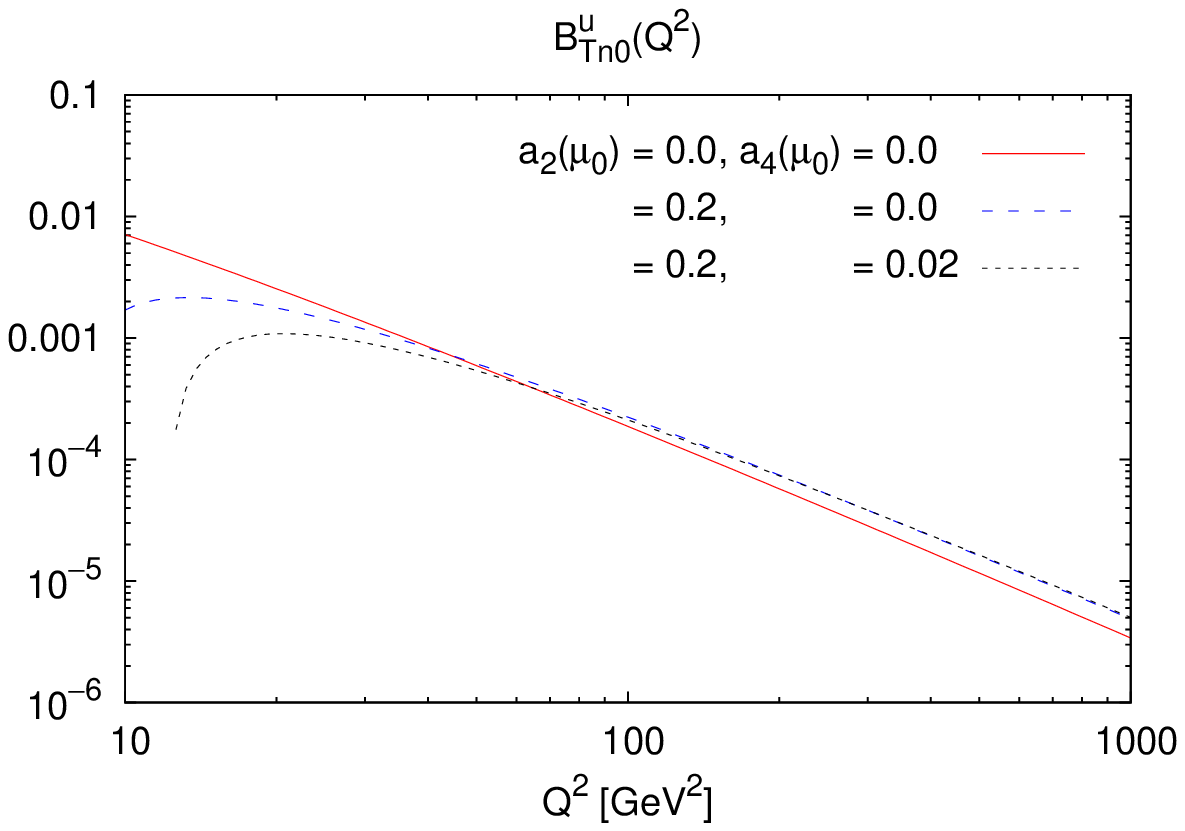}
\end{center}
\caption{\label{fig:Bn0} The result \protect\eqref{master-result} for the
  form factor $B_{T n0}^u$.  Unless specified in the figure keys, we set
  the renormalization and factorization scale to $\mu^2 = \Lambda Q$ with
  $\Lambda = 500 \mev$.  As a default we take all Gegenbauer coefficients
  $a_n$ to be zero; the reference scale for nonzero values of $a_n$ is
  $\mu_0 = 2\gev$.}
\end{figure}

Let us now take a look at our result \eqref{master-result} for $B_{T
  n0}^u$.  Since the loop integral in \eqref{ff-kperp} receives
contributions from gluon virtualities ranging all the way from order $Q^2$
to order $\Lambda^2$, an adequate choice for the renormalization and
factorization scales may be to take the geometric mean $\mu^2 = \Lambda
Q$, which we take as a default in the following.  In the first panel of
Fig.~\ref{fig:Bn0} we compare the results obtained with this choice and
with the naive choice $\mu^2=Q^2$.  The differences are noticeable but not
as large as the ones we discuss next.

In the second panel of Fig.~\ref{fig:Bn0} we compare the form factor
calculated with three different values of the effective parameter
$\Lambda$, where the central value $\Lambda = 500 \mev$ corresponds to an
estimate based on a model of the pion wave function \cite{Jakob:1993iw},
as discussed in the appendix.

In the third panel of the figure we investigate the sensitivity of our
result to the twist-two pion distribution amplitude.  The difference
between the three example choices for the lowest two Gegenbauer
coefficients are quite small at high $Q^2$ but very noticeable as $Q^2$
decreases.  We note that the two curves with $a_2(\mu_0) = 0.2$ have a
zero crossing, which occurs at $Q^2 = 7.8\gev^2$ for $a_4(\mu_0) =
0$ and at $Q^2 = 12.2\gev^2$ for $a_4(\mu_0) = 0.02$.  This behavior
can be understood from \eqref{master-result}.  Compared with the term
proportional to $\ln^2 Q/\Lambda$, the contribution linear in $\ln
Q/\Lambda$ has a global minus sign and larger numerical coefficients
multiplying the $a_n$.  If $\ln Q/\Lambda$ is not large enough, the linear
term can therefore dominate and give a negative result for positive $a_n$.
As we discussed after \eqref{master-result}, the strong enhancement of
contributions from higher $a_n$ is to taken with great caution, and we
therefore do not regard the occurrence of a zero crossing for $B_{T n0}^u$
as a reliable prediction.

We note that all curves in Fig.~\ref{fig:Bn0} fall less steeply than a
pure power law $1/Q^4$.  This is to be expected since the enhancement by
the squared logarithm of $Q^2/\Lambda^2$ is stronger than the decrease
from the scale dependence of $\alpha_s(\mu)\, \mu_\pi(\mu) \sim
\alpha_s(\mu)^{\, 0.52}$.

Let us finally compare the different form factors for our default choices
$\mu^2 = \Lambda Q$ with $\Lambda = 500 \mev$ and $a_n=0$.  The ratio
$B_{T n0}^u /B_{T 32}^u$ varies between 35 and 240 for $Q^2$ between $10$
and $1000\gev^2$.  At $Q^2 = 10\gev^2$ we find that $B_{T n0}^u$ is about
two thirds of $F_\pi$.  It is amusing that we obtain $B_{T n0}^u = 0.038$
at $Q^2 = 2.5\gev^2$, which is within a factor of a few from the results
obtained for $B_{T 10}^u$ and $B_{T 20}^u$ in the lattice calculation
\cite{Brommel:2007xd}.  This coincidence must, however, not be
over-interpreted, given the uncertainties we have just discussed and given
that we have not evaluated the $\mathcal{O}(1)$ contribution in
\eqref{master-result}, which is different for different $n$ in $B_{T
  n0}^u$.

%%%%%%%%%%%%%%%%%%%%%%%%%%%%%%%%%%%%%%%%%%%%%%%%%%%%%%%%%%%%%

\section{Summary}
\label{sec:summary}

We have studied the tensor form factors of the pion at large squared
momentum transfer $Q^2$.  The matrix element of the chiral-odd quark
currents with twist two are written as the convolution of a
hard-scattering kernel, the twist-two distribution amplitude for one pion
and the twist-three distribution amplitudes for the other pion.  In the
twist-three sector we take the asymptotic form of the two-particle
distribution amplitudes, so that the three-particle distribution
amplitudes do not contribute \cite{Braun:1989iv,Beneke:2000wa}.

For the $\xi$-dependent part of the matrix element \eqref{ff-def}, i.e.\
for the form factors $B_{T ni}^u$ with $i\ge 2$, one can take the
collinear limit of the hard-scattering kernel.  The result is a
representation in standard collinear factorization, in full analogy with
the well-known expression \eqref{fpi} for the electromagnetic pion form
factor $F_\pi$.  The form factors $B_{T ni}^u$ with $i\ge 2$ behave like
$1/Q^4$ up to logarithms from the scale dependence of $\alpha_s$ and
$\mu_\pi^{} = m_\pi^2 /(m_u+m_d)$.  Numerically, we find that $B_{T 32}^u$
is more than a factor 10 smaller than $F_\pi$ already at $Q^2 = 5\gev^2$.

For the form factors $B_{T n0}^u$ the collinear limit cannot be taken,
because the hard-scattering formula then develops logarithmic divergences
in the integrations over the longitudinal momentum fraction of the quark
in both the incoming and outgoing pion.  We have used a simple
regularization of the collinear divergences, which involves an effective
parameter $\Lambda$ representing the typical transverse momentum in the
gluon propagator of the graphs in Fig.~\ref{fig:graphs}.  The momentum
fraction integrals then give enhancement factors $\ln^2 Q/\Lambda$ and
$\ln Q/\Lambda$ that modify the $1/Q^4$ power behavior of $B_{T n0}^u$.
This is reminiscent of the analysis in \cite{Belitsky:2002kj}, where the
$1/Q^6$ power behavior of the proton Pauli form factor $F_2(Q^2)$ was
found to be modified by a squared logarithm $\ln^2 Q/\Lambda$ related with
end-point divergences in a purely collinear calculation.

We have evaluated the logarithmically enhanced terms for $B_{T n0}^u(Q^2)$
and find that they are independent of the moment index $n$.  These terms
depend very strongly on the end-point behavior of the twist-two
distribution amplitude $\phi(u)$, or equivalently on the Gegenbauer
coefficients $a_n$ with high $n$. We expect this dependence to be
decreased by Sudakov effects, which suppress the end-points at
sufficiently large $Q^2$.  Numerically, we find that for $Q^2 > 10\gev^2$
our approximation of $B_{T n0}^u$ is considerably larger than $B_{T ni}^u$
with $i\ge 2$, which is a direct consequence of the enhancement factor
$\ln^2 Q/\Lambda$.

In the present work we have deduced the basic behavior of the form factors
$B_{T ni}^u$ at large $Q^2$.  An evaluation that could claim to be
quantitatively valid at moderately large $Q^2$ would need to use a
formalism with a more realistic treatment of the end-point regions in the
momentum fractions.  Obvious candidates for this are the modified
hard-scattering formalism
\cite{Li:1992nu,Jakob:1993iw,Goloskokov:2009ia,Li:2009pr} or approaches
based on QCD sum rules
\cite{Nesterenko:1982gc,Braun:1999uj,Bakulev:2004cu}.

%%%%%%%%%%%%%%%%%%%%%%%%%%%%%%%%%%%%%%%%%%%%%%%%%%%%%%%%%%%%%

\section*{Acknowledgments}

We are grateful to V.~Braun and Th.~Feldmann for helpful conversations.
Special thanks go to B.~Pire for numerous discussions and advice.

This work was supported by the exchange program PROCOPE of the German
Academic Exchange Service and the French Minist\`ere des Affaires
\'Etrang\`eres.
L.Sz.\ is partially supported by the Polish grant MNiSW N202 249235.  He
also acknowledges the warm hospitality at CPhT of \'Ecole Polytechnique
and at LPT in Orsay.

%%%%%%%%%%%%%%%%%%%%%%%%%%%%%%%%%%%%%%%%%%%%%%%%%%%%%%%%%%%%%

\appendix

\section{A simple estimate of $\Lambda$}
\label{sec:appendix}

In order to get some feeling for the typical size of the effective
parameter $\Lambda$, let us take a closer look at the replacement of
$(\tvec{k}-\tvec{k}')^2$ by $\Lambda^2$ in \eqref{k-to-lambda}.  To this
end we assume that $\Sigma(u,k^2)$ is independent of $u$, so that we can
still perform the integrations over $v$ and $u$ as in \eqref{ff-lambda} to
\eqref{master-result}.  The logarithms $\bigl[ \ln (Q^2/\Lambda^2)
\bigr]^p$ with $p=1,2$ in \eqref{master-result} should then be replaced by
\begin{align}
  \label{kt-integrals}
\int d^2 k\, d^2k'\; \Sigma(k^2)\, \Sigma(k'^2)\,
  \biggl[ \ln \frac{Q^2}{(\tvec{k}-\tvec{k}')^2} \biggr]^p .
\end{align}
Let us for simplicity assume a Gaussian form
\begin{equation}
  \label{gauss-wf}
\Sigma(k^2) = \frac{1}{2\pi \sigma^2}
   \exp\biggl[ - \frac{\tvec{k}^2}{2\sigma^2} \biggr] ,
\end{equation}
where $\sigma^2$ is the average squared transverse momentum in the pion
wave function.  In a study of $F_\pi$ using the modified hard-scattering
picture of Li and Sterman, this parameter has been estimated as $\sigma
\approx 350 \mev$ in conjunction with the twist-two distribution amplitude
$\phi(u) = 6 u\bar{u}$ \cite{Jakob:1993iw}.

With \eqref{gauss-wf} one can readily perform the integrals
\eqref{kt-integrals} after a change of variables from $k$ and $k'$ to
$k+k'$ and $k-k'$.  The result is
\begin{align}
  \label{integrals-done}
& \frac{1}{(2\pi)^2\ms \sigma^4} \int d^2 k\, d^2k'\;
  \exp\biggl[ - \frac{\tvec{k}^2 + \tvec{k}'^2}{2\sigma^2} \biggr]
  \ln \frac{Q^2}{(\tvec{k}-\tvec{k}')^2}
\nonumber \\
&\quad = \ln \frac{Q^2}{4 e^{-\gamma}\ms \sigma^2} \, ,
\nonumber \\[0.2em]
& \frac{1}{(2\pi)^2\ms \sigma^4} \int d^2 k\, d^2k'\; 
  \exp\biggl[ - \frac{\tvec{k}^2 + \tvec{k}'^2}{2\sigma^2} \biggr]\,
  \biggl( \ln \frac{Q^2}{(\tvec{k}-\tvec{k}')^2} \biggr)^2
\nonumber \\
&\quad = \biggl( \ln \frac{Q^2}{4 e^{-\gamma}\ms \sigma^2} \biggr)^2
   + \frac{\pi^2}{6} \, ,
\end{align}
where $\gamma = -\int_0^\infty dx\, e^{-x}\ms \ln x$ is Euler's constant.
The term $\pi^2/6$ can be neglected in our approximation, so that we can
consistently identify the first and the second expression in
\eqref{integrals-done} with $\ln (Q^2/\Lambda^2)$ and $\ln^2
(Q^2/\Lambda^2)$, respectively.  We thus find that with the
transverse-momentum dependence \eqref{gauss-wf} of the pion wave function
we have $\Lambda = 2 e^{-\gamma/2}\ms \sigma \approx 1.5\ms \sigma$, which
according to the above estimate for $\sigma$ corresponds to $\Lambda
\approx 525 \mev$.

%%%%%%%%%%%%%%%%%%%%%%%%%%%%%%%%%%%%%%%%%%%%%%%%%%%%%%%%%%%%%

\end{document}